\documentclass{elsart}
\usepackage{amssymb,amscd,graphicx}
\begin{document}
\runauthor{Gadomski, ETAL}
\begin{frontmatter}

\title{On temperature- and space-dimension dependent matter agglomerations in a
mature growing stage}
\author{A. Gadomski$^a$},
\author{J.~M. Rub\'{\i}$^b$\corauthref{cor}},
\author{J. \L uczka$^c$},
\author{M. Ausloos$^d$}
\address{
$^a$Institute of Mathematics and Physics,
  University of Technology and Agriculture,
PL--85796 Bydgoszcz, Poland} 
\address{$^b$Departament de F\'{\i}sica Fonamental, Universitat de Barcelona, 
E--08028 Barcelona, Spain} 
\address{$^c$ Institute of Physics, University of Silesia,
PL--40007 Katowice, Poland} 
\address{$^d${\em SUPRATECS}, University of Li$\grave e$ge,
B-4000 Li$\grave e$ge, Belgium}

\corauth[cor]{Corresponding author, fax: +34 934021149, e-mail:
mrubi@ffn.ub.edu}

\date{}
\maketitle
\begin{abstract}
Model matter agglomerations, with temperature as leading control parameter, have been considered, and some of their characteristics have been studied.
The primary interest has been focused on the grain volume fluctuations, the magnitude of which readily differentiates between two commonly encountered  types of matter agglomeration/aggregation processes,
observed roughly for high- and low-density matter organizations.
The two distinguished types of  matter arrangements have been described through the (entropic) potential driving the system.
The impact of the potential type on the character of matter agglomeration has been studied, preferentially for (low-density) matter aggregation for which a logarithmic measure of its speed has been proposed.
A common matter diffusion  as well as mechanical relaxation picture, emerging during the mature growing stage, has been drawn using a phenomenological line of argumentation. Applications, mostly towards obtaining soft agglomerates or so--called jammed materials, have been mentioned.
\end{abstract}

\begin{keyword}
soft matter agglomeration and aggregation; fluctuations; temperature; space dimension; power law; entropic potential; diffusion; drift; relaxation \\
{\PACS 81.10.Jt, 82.70.Rr, 64.60.-i, 05.40.-a}
\end{keyword}
\end{frontmatter}

\section{Introduction}

\indent Agglomeration and aggregation of matter \cite{kaye} are ubiquitous phenomena expected to occur almost everywhere in nature. Their appearances range from metallic polycrystals to colloidal aggregates and clusters as well as lipid-protein viscoelastic matrices called often biomembranes.
As observed in colloids and protein aggregates immersed in solutions \cite{henk,derjag,vekilov} and in granular media, such as (wet) sands \cite{fin}, we can basically expect two types
of emerging aggregations, namely, those termed diffusion-controlled, for which the attachment of particles and clusters occurs instantaneously at first touch, and the others, which are chemical-reaction controlled. For the latter, the 'first touch' sticking principle does not hold, and the microstructural evolutions have to follow certain attachment-detachment sequences of events.
It finally results in expected matter incorporation in the form of a particle,  or some cluster of particles, being attached to a colloidal floccule or protein aggregate, for example.
The resulting microstructures, however, are of different types \cite{henk,fin}. Those, which are   chemical-reaction controlled are rather more compact, whereas their diffusive counterparts look more irregular {\it viz} fractal, meaning that their arms or branches expand more visibly, so that they are quite undense as a whole \cite{jullien,turnbull}.
A qualitatively similar observation can be offered while looking  at matter agglomeration in some systems investigated in metallurgy \cite{christian}. Although one can find many qualitative similarities in the emerging structures, there are differences mostly due to different types of the matter fluctuations, so that one could be able to distinguish among the emerging transformations \cite{dk,christian}.

We may easily observe that there are matter agglomerations that form typically
closely-packed (CP)  microstructures. Polycrystals, ferroelectrics or ferromagnets at sufficiently low temperatures, {\it i.e.} composed of electric or magnetic domains, respectively, fall into this category \cite{vandew}. Binary alloys may also build structures in a domainwise way. In such systems, the total volume is a conserved quantity \cite{npcs}.
On the contrary, there exist processes, such as phase separations in solid solutions,
as for example the sintering of powders \cite{kaye}, that lead to the formation of relatively loosely-packed (LP) microstructures such as immersion-containing assemblies, or certain composites \cite{turnbull}.
These certainly do not occur under the constraint of constant total volume, see  \cite{physica1}, and Ref. [25] therein.

The potential governing the system, {\it viz} acting on the grains, plays an important role in a mesoscopic description. It has been shown for a type of model grain growth processes, that a logarithmic potential emerges, which is associated with the  pivotal role played by the curvature in such processes. The result obtained seems to be valid for many of the agglomerations whose emergence obeys the Laplace-Kelvin-Young law \cite{acta}. The results are also of importance for soap froths, foams  and typical bubble-containing cellular microstructures,  droplets and microemulsions as well. All together they constitute the agglomerates that we name CP microstructures \cite{physica2}. This seems to be in contrast to LP microstructures or aggregates. Based on the mesoscopic nonequilibrium thermodynamic description, in the present paper we are following an  approach, that leads to a power form of the potential driving the system. This power form differs, however, from another genuine power form that we have derived previously and referred to as Lennard-Jones form throughout \cite{chemphys}. In fact, in an Euclidean space, in which the distance is defined by the grain radius, it is a linear form. Otherwise, it has algebraic form.

The basic aim of the present study is to propose a model suitable to reveal essential differences between CP and LP types of matter agglomerates \cite{physica2,physica1}. Since in \cite{chemphys,physica2}, and in other studies \cite{npcs,physica3}, CP agglomerations have been studied quite exhaustively, in the present study we wish to place further emphasis on LP agglomeration, describing it in detail.
Moreover, we would like to offer a comprehensive picture of the matter agglomeration {\it versus} aggregation effects, understood roughly in terms of a sol-gel phase change, focusing the attention, however,  not on the type of the phase transition but on the physical factors controlling the phenomena in question. These are: temperature, space dimension as well as a type of interactions, but taken exclusively at the cluster level.
 To get an overview of the phenomena, we are also interested in viscoelastic properties of the slowly growing agglomerate in its final stage of evolution.

A special motivation for the present study appears to be, that in the preceding studies \cite{npcs,acta,physica1,physica3} we derived the same growth rule for both CP and LP agglomerations. We found, however, certain differences between both processes: They do not behave in the same way as far as  both, the number of grains  as well as the total volume, are concerned \cite{npcs}. Now,
 we wish to see which are the most essential differences between agglomerations grain volume fluctuations ${\sigma }^2 (t)$.
 These show a different behavior because matter organization during agglomeration is different 
\cite{earn} under local temperature-damage conditions, as was also observed in fracture phenomena \cite{maus,sorn}.
It seems to be almost exactly like in the thermal fuse model, designed for fracture phenomena, in which the same quantities, such as temperature and strain-stress factor play a leading role \cite{sorn}.

The plan of the paper is as follows. In Section 2, we propose a certain summary of the principal characteristics of the two agglomerations/aggregations
processes. We emphasize a simple derivation of a new (entropic) potential form 
and point to a similarity between two types of random walks. In Secion 3,  we consider
 the grain volume fluctuations. In Secion 4, we discuss the overall context of the late-time growing phenomena of interest, coupled additionally to mechanical stress relaxation, observed during late stage of growth. Conclusions are in Section 5.

\section{CP and LP agglomeration processes and their characteristics}

\indent In two recent studies \cite{physica2,chemphys} in which we have worked out a thermodynamic-kinetic description of model complex matter CP agglomerations, as well as in \cite{physica1}, where LP agglomerations (aggregations) have been considered, we have used the expression for the current in the space of cluster volumes \cite{physica2} 
\begin{eqnarray}
\label{JJ}
J (v,t) = - D v^{\alpha} {\partial \over \partial v}  f  - \Big[b(v) {\partial \over \partial v}
\phi\Big] f,
\end{eqnarray}
where $f\equiv f(v,t)$ stands for the number density of grains of volume $v$, which means, that $f (v,t) dv$ is a (relative) number of grains of a size taken from the narrow volume interval $[v,v+dv]$ \cite{acta,npcs}; $t$ is the time. $\phi $ represents the physical potential (see, \cite{chemphys} for an explanation of the term), assumed to be one of the  relevant drivers for the agglomeration process at the mesoscopic level, and assuring its nonequilibrium character \cite{physica2,chemphys}. Moreover \cite{physica2},
\begin{eqnarray}   \label{bv}
b(v) = {D\over {k_B T}} {v^\alpha }, \qquad \alpha = {1 - {1\over d}},
\end{eqnarray}
where $D$ is a diffusion reference constant; note that a principal role of $D$ is to scale the time variable. $k_B$ represents the Boltzmann constant.
Note that the mobility $b(v)$ is related with the Onsager coefficient, $L(v)$, that appears in derivation of the matter flux (\ref{JJ}) under a set of assumptions, mostly on locality of the process, {\it etc.}, for details see \cite{physica2}.
$L(v)$, and hence $b(v)$, could be measured 
by comparing the current and the thermodynamic force.
The quantity $D v^\alpha $ is to be inferred from the Green-Kubo formula, so that there would be quite  a strong theoretical support for deriving $b(v)$ both, experimentally as well as theoretically \cite{physica2}.
In  \cite{physica2,chemphys} we have derived a (so--called) compaction potential
\begin{eqnarray}   \label{philn}
\phi (v) = \phi_o ln{({v/v_o})}, 
\end{eqnarray}
where $\phi _o$, $v_o$ - constants. While speaking about the role of the logarithmic potential, we wanted to see it as  a quantity  assuring emergence of rather compact
 structures \cite{physica2,npcs}.
Its role  is equivalent to the action of grain curvatures \cite{physica2}, resulting
for a given value of the intergrain surface tension,  in emergence of the driving force, such as
the one involved in the Laplace-Kelvin-Young law \cite{acta}.

In a previous study on the phase transformation kinetics for LP "diffusive" agglomerates we have started with the matter flux of a purely diffusive nature \cite{physica1,npcs}, namely
\begin{eqnarray}
\label{JD}
J (v,t) = - D v^{\alpha} {\partial \over \partial v}  f.
\end{eqnarray}
The diffusion function  $D(v) = {D v^{\alpha }}$
is  proportional to the grain surface.

Both CP and LP processes follow from the general form (\ref{JJ}).
Indeed, the case of LP is  obtained
 when the second term in r.h.s. of  (\ref{JJ}) can be neglected.
Formally, $b(v) \to 0$ when $T\to \infty$. From the physical point of view, it corresponds to sufficiently high temperatures  $T \ge T_{th} > 0$,
where $T_{th}$ can be treated as  a threshold temperature above which
 the agglomeration  takes place exclusively by yielding LP microstructures.
However, the second term in r.h.s. of  (\ref{JJ}) depends both on $v$ and $T$.
It tends {\it uniformly} (i.e. independently of $v$) to zero at the
high temperature limit  if
\begin{eqnarray}   \label{cond1}
b(v){{\partial \phi (v)}\over{\partial v}} = const.
\end{eqnarray}
Then for a given system, temperature $T_{th}$ does not depend on $v$
and  is consistently defined.   It is a case when the potential
\begin{eqnarray}   \label{phiv}
\phi (v) \propto v^{1 - \alpha } = v^{1/d}.
\end{eqnarray}
It should be noted that $\phi $ is an entropic potential \cite{chemphys} which can be a cause of some disaggregation, or at least, cessation-to-growth (or, impingement) effects, taking place within the overall aggregation space.

The above potential form (\ref{phiv}), designed for LP agglomeration, seems convincing as well as legitimate here: note that the 'force'
\begin{eqnarray}   \label{fgg}
{F_{g-g}} \propto {{\partial \phi (v)}\over {\partial v}}
    \propto {1\over v^\alpha} \propto {1\over R^{d - 1}},
\end{eqnarray}
%
%
%
because  $v\sim R^d$, with $R$ - the single grain radius. Thus, $F_{g-g}$ acts as the inverse of the area of the grain hypersurface which implies that the smaller the area is, the bigger the force acting on the grain can be, thus  impeding the formation of new grains, which would contribute to an agglomerate's density increase. Note that this way we have established a certain matter densification procedure, expected to work in the systems of interest.
Qualitatively, the similar dependence is for the CP matter aggregation:
from (\ref{philn}) one gets for the 'force'
\begin{eqnarray}   \label{ffgg}
{F_{g-g}}     \propto {1\over v} \propto {1\over R^d}.
\end{eqnarray}
Here, $F_{g-g}$ acts as the inverse of the hypervolume of the grain, which makes a difference between CP and LP agglomeration. 
Thinking in a somewhat different way, some relaxation of the surface tension conditions \cite{physica2} is expected to occur in the LP case, and some visible escape from the Laplace-Kelvin-Young strategy, {\it cf.} the review \cite{npcs} for details, characteristic of polycrystals and cellular microstructures \cite{acta} can be  foreseen. Moreover, the mean grain curvature may not be necessarily a dominating physical factor: either another type of curvature, the Gaussian curvature, may prevail \cite{kash}, and the so-called model nano-agglomeration may occur \cite{chemphys}; or any relevant signatures of curvature contributions, as important thermodynamic-kinetic factors accelerating agglomeration, are not found at all \cite{physica1}, as occurs in some dispersive systems \cite{derjag,israel}.  This is also what we mean by the relaxation of the surface tension conditions for loosely-packed grain systems.

Referring further to (\ref{cond1}) and using as in \cite{chemphys,npcs}
the similarity relation, $v\sim R^d$, one gets
\begin{eqnarray}   \label{phiR}
\phi (R) \propto  {\phi ({R_o})} {R\over R_o},
\end{eqnarray}
where
\begin{eqnarray}   \label{phiRo}
{\phi ({R_o})} = {k_B T\over {D_{\alpha }}} R_o, \qquad T \ge
 T_{th} ,
\end{eqnarray}
and, in consequence, $\phi (R) \propto R$; ${D_{\alpha }} = D {(1 - \alpha )}$,  $R_o$ as the initially observed grain radius is usually of the order of a few $\mu m$, or somewhat bigger \cite{chemphys,physica2}, which implies a typical length scale of colloids \cite{derjag}. Notice, that while exploring the similarity relation\footnote{As an extension of $v\sim R^d$, one could think of clusters of fractal dimensions
$0 < d_f < d$, aggregating in an Euclidean $d$-dimensional space: It is interesting to note that
such cluster-cluster (colloidal) aggregation studies, with their structural as well as kinetic aspects involved, are mostly performed in $d=2$, and for some of them, $d_f \approx 1.6$ was measured by several different methods \cite{earn}. Note that the potential $\phi (R)$ from (\ref{phiR}) will no longer be linear in $R$ when a modification like
$v\sim R^{d_f}, 0 < d_f < d$ would finally be used. On the contrary, it will take an algebraic (nonlinear) power form in $R$, with a characteristic exponent of $0 < {d_f /d} <1$.
}
 $v\sim R^d$, one immediately sees that the state-dependent (grain-surface related) diffusion coefficient $D(v) \simeq D(R^d ) \propto R^{d-1}$ is always obtained. Let us recall right here that for certain model biomembranes, that can be both reactive as well as constrained curvilinear systems,  a modified kinetic coefficient, like $D(v,t) = D {v^\alpha } (t + t_{ch})^{-h}$, with $t_{ch}$  a characteristic time, and $h$  a microstructure-sensitive exponent \cite{npcs}, has also been proposed.
The relation obtained can be anticipated as characteristic of systems with well-developed surfaces.

Some supporting explanation of the surface-tension  relaxation effect, from which the LP agglomeration may readily result, is under CP-agglomeration conditions. When  the grains are  packed densely together, the boundary of an individual grain can quite strictly be equivalent to a corresponding piece of the typically very thin overall intergrain separation space of the agglomerate. When the high-temperature activation conditions are switched on, the intergrain separation space would tend to increase. This means that now the grain boundary is not as before, but very likely a second diffusive layer emerges, which  effectively thickens its total width. Such a scenario is often associated with the double layer (DL) of Stern type, and bears a visible signature of phase separation effect \cite{adam}.


Now, let us try out a phenomenology, supporting a plausible passage between the random walk in the  space of cluster volumes \cite{npcs,acta}, and the random walk, that is typically observed when studying the cluster-cluster phenomena, {\it i.e.} that of Einstein-Smoluchowski (ES) type \cite{jullien}. In contrast to the agglomeration/aggregation phenomena  that we systematically propose to consider \cite{physica1,physica2} in the space of grain/cluster sizes, the ES random walk is realized in the space of positions of the centers of mass of moving clusters \cite{henk}. 
For further remarks on a certain equivalence between both types random walks, and which are the (fractal) measures of it, see \cite{physica3}. 
Why would one speak of such an equivalence ? Because both random walks in question can reasonably be anticipated as 'doing the same', at least in a thermodynamic sense: A (local) move of the grain or cluster boundary forward, associated with a step over an energetic barrier would be equivalent to an incorporation of the particle (macroion; molecular cluster), that, in turn, must overcome a thermal energy barrier to be a part of the boundary, or to be electrostatically caught by it. Otherwise, the boundary fluctuates, experiencing no advancement in the overall hypervolume space, so the particle also moves, performing its random walk. This is a typical scenario in a DL of Stern type and coincides well with cluster-cluster ES like picture \cite{adam,kisza,physica3}.

If we accept an ES-like picture of the cluster-cluster aggregation process, completed by the involvement of DL, we can arrive at the following quantitative description while approaching the temperature threshold $T_{th}$.
As can be learned from \cite{coniglio}, where a complex viscoelastic behavior at the sol-gel transition has been sketched as $T\to {T_{th}}$ from below, a sol phase, typically containing small clusters, the average linear sizes of which diverge at the mentioned temperature limit, merges rapidly into a bigger (gel) mega-cluster. Under such circumstances, according to \cite{coniglio} (see, Eq. (1) therein), the solvent viscosity, $\eta _s$, involved in the  Einstein-Stokes (E-S) relation, $D_\alpha = {k_B T}/{6 \pi {\eta _s} R_H}$ \cite{talkner,adam}, where $R_H$ stands for the hydrodynamic radius of the cluster, becomes cluster-radius dependent which explains the coagulation phenomena, often assisted by some changes in solution's viscosity \cite{martin}.
Such a possibility can easily be inferred from our modeling as well. 

Let us present it for spherical clusters. When applying formula (\ref{phiRo}) to the E-S formula,
one is able, upon identifying $R_H$ with $R_o$, to write
\begin{eqnarray}   \label{est}
{D_\alpha } = {{k_B T}\over {6\pi {\eta _s} R_o}},
\end{eqnarray}
where, according to slow growth requirement postulated,  $\mid R - R_o \mid$ would always tend to a small value for almost all $t$-s taken from the time zone of interest \cite{acta}, thus enabling to make such an identification; $R\equiv R(t)$.  In  consequence,  an expression for the solvent viscosity can be proposed
\begin{eqnarray}   \label{solvent}
\eta _s = {{\phi (R_o )}\over {6\pi {R_o}^2}} ,
\end{eqnarray}
as well as the characteristic time $\tau   = 1/\eta _s$
\begin{eqnarray}   \label{time_ch}
{\tau  }= {{3 a_o} \over {2 \phi(R_o)}},
\end{eqnarray}
where $a_o = 4\pi {R_o}^2$. 

By presenting the above picture we would like to point out a striking analogy between the sol-gel transition, and the LP-CP transition: Note that to accept the analogy, the temperature course has to be inverted here because, according to our previous explanations, the CP phase is usually a low temperature phase, and in consequence, LP emerges at higher temperatures. 

As was indicated in \cite{coniglio}, the E-S diffusion
coefficient depends on $R_o$ but not on temperature, when the system is at the critical regime. The above
described situation would invoke recent studies on the so-called
kinetic arrest during gel formation, which can halt phase
separation due to spinodal decomposition \cite{weitz}, thus
enabling the gel formation. If the kinetic arrest is really the
case, two physical quantities appear to be of relevance: the space dimension and
the magnitude of the volume fraction {\it viz} the
aggregate density become important \cite{jullien2}, 
apparently at the expense of domination of temperature, which is
normally a prevailing physical factor \cite{rubivil,dawson}. A
subtle interplay of attraction-repulsion fields \cite{weitz}, due
to available electrostatics is a key phenomenon, so that our
electrostatics-oriented picture of the percolative-in-nature
process \cite{jullien2} is also supported this way. Such a distinction between 
reaction and non-reaction, {\it viz}
diffusion limits is also very characteristic of colloidal (or, alike, such as those protein-involving \cite{vekilov})  
aggregations, quite irrespective of the space dimension $d$
\cite{earn,adam,henk}. 
\section{The grain volume fluctuations in the late-time growing stage of the agglomeration process}

\noindent
There seems to be no doubt \cite{christian,turnbull,derjag,henk} that both agglomerations in question emerge in a fluctuating matter environment. Therefore, any reasonable quantification of the so-called fluctuation impact on the speed of the matter evolutions under study is worth examining here. Thus, we propose below an evaluation of the reduced variance
\begin{eqnarray}       \label{frf}
{{\sigma }^2 (t)} = {{<v^2 (t)> - {<v^1 (t)>}^2}\over {{<v^1 (t)>}^2}}\equiv  {{<v^2 (t)>\over {<v^1 (t)>}^2} - 1} ,
\end{eqnarray}
as a direct measure of the grain volume fluctuations.

The notation used in Eq. (\ref{frf}) refers to the statistical moments
\begin{eqnarray}       \label{mom}
{<v^n (t)>} = \int_0^\infty v^n f(v,t) dv \qquad  n = 0,1,2, ...
\end{eqnarray}
of the agglomeration process, where the process is usually described by a local continuity equation 
\begin{eqnarray}       \label{loccont}
{{\partial f (v,t)}\over {{\partial t}}} - {{\partial J (v,t)}\over {{\partial v}}}, 
\end{eqnarray}
supplemented by the corresponding initial (of delta-Dirac type  as a first successful attempt) and boundary (typically, of absorbing type) conditions (IBCs)
 \cite{acta,physica1,npcs,physica3}.
The explicit solutions, $f(v,t)$--s, are presented elsewhere \cite{physica1,acta,npcs}.

Then zeroth moment, $<v^0 (t)>$, is related to the average number of grains in the system, and usually  suffers from an algebraic drop with time \cite{physica1,acta}. The first moment, $<v^1 (t)>$, is related to its total volume
which is a constant value for CP agglomerations \cite{acta,npcs} and an increasing function of time for LP agglomerations \cite{physica1}. From the expressions of  both moments, it follows that the average grain radius, $R_{av} (t)$, behaves powerly with time, with
growth exponent $1/(d+1)$ that apparently carries some signature of random close-packing of matter by having included the super-dimension $d+1$ \cite{physica3,physica1,physica2,chemphys}. These constitute the main characteristics of the model agglomeration/aggregation  process in its late-stage ($t>>1$) limit.

The question remains about asymptotic values of the moments $<v^n (t)>$ that must be known when applying formula (\ref{frf}). For CP agglomerations, they are found to obey a power law \cite{acta,npcs}
\begin{eqnarray}       \label{frfCP}
{<v^n (t)>} \sim t^{(n-1)/(2-\alpha )} \qquad (n=0,1,2), \quad t >> 1 ,
\end{eqnarray}
whereas for LP agglomerations one provides another power law \cite{physica1,npcs}
\begin{eqnarray}       \label{frfLP}
{<v^n (t)> } \sim t^{[(n-1)+\alpha ]/(2-\alpha )} \qquad (n=0,1,2), \quad t >> 1.
\end{eqnarray}
Notice, that for $\alpha = 0$ ($d=1$) both power laws above converge to the same form, namely ${<v^n (t)>} \sim t^{(n-1)/2}$.
When applying (\ref{frf}) and (\ref{frfCP}) it appears that for CP agglomerations,  ${{\sigma }^2 (t)}$ can be fully identified with the inverse of $<v^0 (t)>$ (the average number of grains), {\it cf.} \cite{acta,npcs} for details, what because of the constancy of $<v^1 (t)>$, leads to ${{\sigma }^2 (t)} \propto v_{sp} (t)$, where ${v_{sp}}\equiv v_{sp} (t) \simeq <v^1 (t)>/<v^0 (t)>$, and can be named the mean specific volume of the CP agglomerate, being  equivalent to the inverse of its mean number density. The specific volume fluctuations, which are very important in nucleation phenomena  \cite{kash,christian}, are
\begin{eqnarray}       \label{frfv}
{{\sigma }^2 (t)} \propto {t^{d/(d+1)}} ,
\end{eqnarray}
and if $d\to\infty$, ${{\sigma }^2 (t)} \simeq v_{sp} \propto t$.

When using (\ref{frf}) and (\ref{frfLP}), however, it turns out that for LP agglomerations ${{\sigma }^2 (t)}$ is a quantity equivalent to the average grain radius $R_{av} (t)$, see \cite{physica1,npcs} too, and they behave in time as
\begin{eqnarray}       \label{frfR}
{{\sigma }^2 (t)} \propto {t^{1/(d+1)}} ,
\end{eqnarray}
and when $d\to\infty$, ${{\sigma }^2 (t)} \simeq R_{av} (t) \to const$.
Note that a diffusional regime, characterized by the one-half exponent, is achieved exclusively for LP agglomerations in $d=1$ because the only (linear)  characteristic is $R_{av}\equiv R_{av} (t)$: Note that $v_{sp} (t)$ is not a linear characteristic, since $v_{sp} (t) \propto {{[R_{av} (t)]}^3}$ usually holds, and here the $d=1$--case must clearly be discarded as 'classically' diffusional, see Eq. (\ref{frfv}) for comparison.

Upon looking more carefully at the formula (\ref{frfv}) and (\ref{frfR}), we have found out an interesting 'harmonic', or efficiency, property for the agglomerations under study. Namely, we have observed, that upon defining a $d$-dependent logarithmic speed for the agglomeration in dimension $d $ as
\begin{eqnarray}       \label{LPsp}
{{\nu _{sp}}^{(d)}} = {ln{[{\sigma }^2 (t)]}\over {ln(t)}} , \quad d = 1,2,3 ,
\end{eqnarray}
we are able, up to a certain but sufficient accuracy level, to see that only for LP agglomerations one is capable of finding the harmonic rule, namely
\begin{eqnarray}       \label{HR}
{2\over {\nu _{sp}}^{(2)}} = {{1\over {\nu _{sp}}^{(1)}} + {1\over {\nu _{sp}}^{(3)}}} ,
\end{eqnarray}
where simply, {\it cf.} Eq. (\ref{frfR})
\begin{eqnarray}       \label{LPsp2}
{{\nu _{sp}}^{(d)}} = {1\over {d+1}}, \quad d = 1-3 .
\end{eqnarray}
Thus, the case of $d=2$ looks as the most efficient from the aggregational point of view, since it contains mean (harmonic) properties of the $d$-dimensional  process subject to its (speed) logarithmic fractal-like measure defined by (\ref{LPsp}).
Many aggregation processes, see \cite{finnis1,finnis2}, where colloidal monolayers are formed on the air-water interface, subjected to many external factors (UV light, AC nd DC currents, ions), and multitude of microstructures emerging in them (such as: fractal, bubble-like, mesomorphic, compact and voids-containing aggregates), have been experimentally observed and quantitatively analyzed, {\it cf.} \cite{earn}, pointing to a Langmuir-Blodgett-like two-dimensional setup as the most frequently used device. For other interesting studies, here on the structure and phase transitions in Langmuir lipid monolayers, see a review \cite{mohwald}.

To sum up in part, let us state that the fluctuations ${{\sigma }^2 (t)}$ have been proposed as a reliable criterion of differentiating between CP and LP agglomerations in dimension $d$, with an emphasis placed on $d=2$, where harmonic-mean properties can be picked up in a natural way. It is known \cite{kt} that the dimension $d=2$, called a critical dimension,  provides a natural framework for order--disorder phase changes in dipolar systems \cite{adam,kisza},  where, in a low-temperature regime, one observes high-energy cost configurations, composed mostly of isolated dipoles, whereas at a high-temperature limit, a plasma (here, a low-energy cost configuration), consisting of some charged entities, emerges. Such a scenario can be compared to coagulation-flocculation effects in colloids \cite{derjag,finnis1,finnis2,earn}, so that the state of high $T$ corresponds to flocculation phenomena - for them our approach with (\ref{phiR}), taken at a mesoscopic level \cite{chemphys}, looks promising. Thus, a LP microstructure, of non-constant hypervolume,  $<v^1 (t)> \ne const$ \cite{physica1}, means a low-energy cost microstructure. Such a differentiation procedure can thus be anticipated as another mesoscopic procedure \cite{rubivil}, helping in the selection of characteristic fluctuations in nonequilibrium instabilities and pattern formation \cite{chohen}, mostly those of hydrodynamic types.

\section{Matter diffusion and stress relaxation}

\indent The above reasoning toward quantifying the fluctuations of the system can be strengthened with a phenomenological co-argumentation. The idea comes from a "coupled" diffusion--relaxation picture that appears in our complex system. In any diffusion-migration growing process, the mechanical strain-stress fields play their role too. In our case, such a situation can be safely expected in the temperature domain $T\le T_{th}$, though another type of relaxation of the stress field, say $\sigma _m$, is expected to prevail when the CP agglomeration conditions are met. A different behavior may be observed when the CP agglomeration conditions are lost for the first time, that is, right at $T= T_{th}$, when the LP context appears. In both temperature regimes, the relaxation of $\sigma _m (t)$ over the course of time, is very likely to go in a way, complementary to what the action of  the matter diffusion-and-migration process, essentially described by the flux (\ref{JJ}). This is expected to occur \cite{physica2,physica1,npcs} presumably under homogeneous strain conditions, $\epsilon _m \approx const$, for $t >> 1$.

An additional motivation for drawing a coupled matter agglomeration and stress relaxation
picture comes from the nature of the agglomeration process which can be considered to  be the
converse of its fracture counterpart. Consider a ($\sigma _m, \epsilon _m$) plane
on which the equilibrium curve is drawn, representing, for a given time instant, the state of a
soft material under stress ($\sigma _m$) that is
elongated as measured by the strain ($\epsilon _m$). After the  quasi-linear
Hooke regime, a nonlinear and plastic regime follows before failure.
One can roughly consider the equilibrium curve to look like a
Van der Waals (cubic equation of state) curve, meaning that the other
extremity  consists of the fragment region, which under pressure
would agglomerate \cite{sorn,maus}.
The region between the minimum and maximum on such a curve is known
to be the siege of a first-order like transition region,
characterized by nucleation and growth \cite{maus}. Moreover, the percolate nature of
fracture process has been evidenced too, and its critical character, associated with the
notion of phase transition of continuous type, with power law characteristics, has also been
emphasized, mostly by numerical modeling of the process \cite{sorn,vandew}.

From \cite{ag_mpl} it can be learned, that in the absence of non-Arrhenius or fractal type kinetics, visibly modifying the diffusion coefficient $D (v)$, one expects the Maxwell dashpot-and-spring model to represent the relaxation behavior.
We wish to set up here a phenomenological picture, showing that both agglomeration and mechanical stress relaxation \cite{ferry}, where the stress relaxation takes place under slow growth conditions, proper of a mature growing stage, are coupled processes. To proceed, we will represent one of the two contiguous and matter-exchanging grains in the agglomerate, say grain $1$, as an expanding one {\it viz} the spring, growing at the expense of its neighbor, to be termed grain $2$ (the dashpot), to which, according to the Maxwell model, the shrinking action should be assigned,
, see \cite{fraction}.

For the system with narrow gaps, the Maxwell's model conditions are almost satisfied, so that the two-grain action can be extended over all pairs of contiguous grains until the expanding (growing) eventually survive. In a next step, the same kind of competition appears as in the well-argumented Laplace-Kelvin-Young scenario suitable for cellular systems \cite{acta}. This picture holds in the CP context.

In the LP context (a system with wide gaps), qualitatively we may have the same picture but with two differences which implies that grain expansion would not be  likely so vigorous that the corresponding gap is wider, therefore untight, so that the fluid leakage might {\it a priori} be more pronounced. Thus, the fluid response against the piston wall is weaker, and the stress relaxation of Maxwell type no longer applies.

The stress relaxation can now be described by introducing a relaxation exponent $\chi $ in the Maxwell-like, or quasi-fractional Maxwell \cite{fraction}, model presented below.  This exponent is, in general, $d$-dependent. Coupled   matter diffusion and stress relaxation processes have been analyzed in \cite{proca} for an anomalous random walk in a position space. Here, we wish to offer a similar picture, but for our random walk performed in the space of cluster volumes.
As is known, the Maxwell stress relaxation picture leads to an exponential decay of the stress:
$\sigma _m (t) \sim exp (-t/\tau  )$, where $\tau  $ is a reference time of the concentrated grain-containing  viscous solution, Eq. (\ref{time_ch}).  This behavior holds for $T < T_{th}$. As mentioned previously, for $T\ge T_{th}$ we propose
\begin{eqnarray}       \label{Max}
{{d\sigma _m}\over {dt}} + {{{\sigma _m}^{\chi }} \over {\tau  }} = 0,
\end{eqnarray}
where the above is usually true when the internal strain field, $\epsilon _m $ is practically constant, which is  the situation expected in rheological model behavior of  colloids and polymers, see \cite{dressler}, and refs. therein. When solving (\ref{Max}), one arrives at
\begin{eqnarray}       \label{Maxs}
{\sigma _m (t)} \sim {{(t/\tau  )}^{-1/(\chi -1)}}, \qquad t >> 1,
\end{eqnarray}
%
where $\chi = 2 d + 3\equiv 2(d+1)+1$\footnote{It is interesting to note here that $\chi = \chi (d)$ can be, upon identifying $d$ as consecutive numbers of (pairwise) emerging bonds in a gelling system, a certain odd-number based generator of the Bethe lattice elements, starting from the 3-bond (initial) generator for $d=0$, and continuing with $d$, what perfectly recovers the number of elements of model Bethe lattice (an ultrametric space), a tool very useful for mean-field description of gels, and other bonds-containing systems \cite{adam,fin,henk}.} The obtained formula, Eq. (\ref{Maxs}), resembles to a large extent an empirical formula ${\sigma _m (t)} \propto t^{-m}$, see Eq. (9.30) in \cite{Van}, characteristic of mechanically relaxing amorphous polymers; a dimensionless exponent $m$ is a measure of the relative importance of elastic and viscous contributions, in our model manifested exclusively by a close relation with $d$, a rough messure of losses in the system {\it cf.} Eq. (\ref{Maxs}). Usually $m$ is proportional to the loss tangent, a well-known dissipation factor in such processes \cite{Van}, so that the energy dissipation is due to the character of the agglomeration (and relaxation) space, characterized by $\chi $ involved in Eq. (\ref{Maxs}). 

The overall relaxation exponent in (\ref{Maxs}) reads
\begin{eqnarray}       \label{x>0}
{1\over {\chi -1}} = {1\over {2(d+1)}},
\end{eqnarray}
which is exactly one half of the growth exponent ${\nu _{sp}}^{(d)}$ given in (\ref{frfR}), see  (\ref{LPsp2}) too. Such a result can be exactly obtained under the assumption of validity of the Hall--Petch (HP) relation \cite{HP}. For $T\approx T_{th}$ \cite{ag_mpl}, one recovers a volume fluctuation-driven growth-and-relaxation synchronous mode 
\begin{eqnarray}       \label{hp}
{\sigma _m} \sim {{R_{av}}^{-1/2}} \sim {\sigma ^{-1}} ,
\end{eqnarray}
which clearly confirms the solution presented above, Eq. (\ref{Maxs}), {\it cf.} relation (\ref{frfR}). Notice that HP relation typically holds for low-temperature systems \cite{HP}, though its range of validity still expands because of its applications to new studied systems \cite{HP,ag_mpl}. Here, in fact, we conjecture a validity of the HP relation outside the low-temperature range, for example in (dense) colloidal gels, or other (reactive) polycrystalline and/or jammed materials (zirconia, that means, ceramics produced by sol-gel method), in which sol-gel characteristics are  very much related to polycrystalline behavior \cite{dorbo,mater}.

Although such a conjectured supposition needs to be proved, it is found to play a significant role in many, mostly plastic as well as superplastic (also, an inverse HP relation,  characteristic of superplastics) metallurgical systems \cite{christian}. It also holds for non-metallurgical systems, such as (bio)polymeric for example, {\it cf.} \cite{ag_mpl,npcs}, and refs. therein.
Proceeding this way, the HP relation couples, at a local thermodynamic equilibrium point near $T_{th}$, diffusion and migration of matter, as well as a complex material relaxation under slow growth in a fluctuating medium, especially in its mature stage, $t >> 1$.

Another relaxation exponent can be defined
\begin{eqnarray}       \label{LPrel}
{{\mu _{sp}}^{(d)}} = {- ln{[{\sigma _m} (t)]}\over {ln(t)}} , \quad d = 1,2,3 ,
\end{eqnarray}
where
\begin{eqnarray}       \label{LPrel2}
{{\mu _{sp}}^{(d)}} = {1\over {2(d+1)}}, \quad d = 1,2,3 ,
\end{eqnarray}
holds, and a very simple relation, like
\begin{eqnarray}       \label{munu}
{2 {\mu _{sp}}^{(d)}} = {{\nu _{sp}}^{(d)}}, \quad d = 1,2,3 ,
\end{eqnarray}
comes out, in fact, as a consequence of the HP condition applied for our complex electrorheological context.

Similarly, the harmonic rules favoring the dimension $d = 2$, in which most of interesting experiments are performed \cite{mohwald}, also apply when looking at ${\mu _{sp}}^{(d)}$, so that the relaxation in dimension two appears to be an expected case and well worth experimenting. Coated materials, recently produced by sol-gel technique deserve recognition, {\it cf.} \cite{mater}, and refs. therein. In consequence, it is believed, that such a complex diffusion--relaxation behavior may readily help in explaining many clustering phenomena taking place under carefully designed experimental conditions, see \cite{finnis1,finnis2,earn} as  selected cases.
In such experiments, an  account of the stress micro-fields, arising in a direct relation with electrostatic fields acting on all the charged objects cannot be underestimated, either at short  or even at long separation distances, see \cite{stamou}. In \cite{finnis1,finnis2} such a micromechanical factor, being a cause of differentiation amongst the obtained latex-sphere-containing microstructures, and the possible influence on fractal or non-fractal properties of the resulting mega-clusters, often leading to different coverage values, has not yet been taken into account.

\section{Conclusions}
The present study is
devoted to a comprehensive picture of the model matter
agglomeration/aggregation phenomena at a mesoscopic (cluster)
level. The integrated picture has been achieved by
offering, mostly in a phenomenological way, some characteristics
that are: temperature, space dimension as well as interaction-type
dependence, and completed by some mechanistic point of view
\cite{ferry}, due to coupled-to-agglomeration relaxation of matter
during late-time growing stage. This can undoubtedly be understood
as a promising new contribution of our study.

The grain
volume fluctuations, generally given by Eq. (\ref{frf}), and in
particular, by Eqs. (\ref{frfv}) and  (\ref{frfR}), for CP and LP
agglomerations, respectively, appear to be a reliable criterion
for understanding the fluctuations in two different types of model
systems, conforming, however, by construction
\cite{physica1,physica2,npcs}, to the same average grain radius involving 
growth rule, namely the
one given by Eq. (\ref{frfR}). It proves then that the growth rule
is not a sufficient criterion to differentiate between two
seemingly different types of processes, even in their mature
stages.

A logarithmic  potential (\ref{philn}) $\phi (R) \propto ln({R/R_o})$
\cite{chemphys}, can also be derived based on an analysis of the
electric component of the interfacial bending moment presented in
\cite{kral} for micelles (cylindrical molecular condensers). A
power-like Lennard-Jones potential, obtained in \cite{chemphys},
also follows from that analysis, and has been derived for micelles
of another type, called spherical condensers \cite{kral,dressler}.
In the present work, we have picked up a potential $\phi $, Eq.
(\ref{phiR}), in its simplest form, {\it i.e.} for non-fractal
objects, being linear in $R$ (and, recognized in elementary
electrostatics as well, but with a clearly specified
{\it viz.} temperature- as well as dimension-dependent prefactor,
$\phi ({R_o})$), Eq. (\ref{phiRo}), certainly  being of entropic
character \cite{rubivil}. 

It is interesting to note that the
harmonic-mean rule, Eq. (\ref{HR}), applied for the logarithmic
speed defined by  (\ref{LPsp}), has been detected: It seems that
any matter compacting CP "procedure" would somehow destroy certain
efficiency conditions, and a natural appearance of spontaneously
formed mesostructures. This may also be the case when not only
$d$-dependent dimensional constraints apply but also when the system is subjected to external
physical factors that may change the number of the degrees of
freedom, causing certain stress-strain fields induced losses in the coverage value, 
{\it cf.} \cite{finnis1,finnis2}, and the pictures therein. 

An overall experimentally-motivated
\cite{kaye,henk,vekilov,christian,ferry} picture that emerges from
our model study shows that the diffusion-migration process may
thoroughly be accompanied by mechanical relaxation, and that
both mutually cooperating sub-processes may go synchronously. 
Note that any approval for
such a synchronous way implies that $\sigma _m$ must also fluctuate,
because $R_{av} $ is identified with $\sigma ^2 (t) $ for a LP
process, so that the quasi-fractional Maxwell model proposed
(\ref{Max}) is likely to work under the requirement that this
quantity fluctuates "accordingly", {\it cf.} Eq. (\ref{hp}), 
{\it i.e.} the overall synchronisation mode. In
consequence, a simple relation, interconnecting both sub-processes
described, may emerge in such a simple form, see (\ref{munu}). 
The
harmonic-mean rule becomes applicable in both coupled cases. Both
processes are recognized as inseparable (recall sol-gel coexisting
phases \cite{mater,coniglio}), and go in the same direction;
notice, the "-" sign in (\ref{LPrel}), which is, however, not the case of
(\ref{LPsp}). Invoking here a comparison presented in the previous
section between agglomeration/compaction and fracture/compaction
processes (being mutually converse), we see also certain
well-established parallels between our model and the scalar
anti-phase mechanical model of rupture with elastic interaction,
{\it cf.} \cite{sorn,maus}.
 
\section*{Acknowledgment}
{\normalsize
One of us (A.G.) would like to mention a support by 2P03B 03225 (2003-2006). Part of M.A.  works is in the framework of an Action de Recherches
Concert\'ee Program of the University of Li$\grave e$ge (ARC 02/07-293)}. We wish to thank Profs. Mike Finnis (Queen's University Belfast), D. Kashchiev (Bulgarian Academy of Sciences, Sofia) and H.J. Herrmann (University of Stuttgart) for their valuable comments on the manuscript.
A.G. also wishes to thank Prof. A. Fuli\'nski, Jagellonian University Cracow,  for his discussion during a seminar lecture held at the Marian Smoluchowski Institute of Physics, and Dr. Tony Paxton, Queen's University Belfast, for drawing his attention to Ref. \cite{christian}.


\end{document}